\documentclass[aps,pra,twocolumn,floats,floatfix,showpacs,amsmath,amssymb,superscriptaddress]{revtex4}

\usepackage{graphicx}
\usepackage{bm}
\usepackage{soul}

\begin{document}

\title{Non-Hamiltonian dynamics in optical microcavities resulting from wave-inspired corrections to geometric optics}

\author{Eduardo G. Altmann}
\affiliation{Max-Planck-Institut f\"ur Physik komplexer Systeme, N\"othnitzer Str. 38, 01187 Dresden, Germany}
\affiliation{Northwestern Institute on Complex Systems, Northwestern University, 60628 Evanston, IL, USA}

\author{Gianluigi Del Magno}
\affiliation{Max-Planck-Institut f\"ur Physik komplexer Systeme, N\"othnitzer Str. 38, 01187 Dresden, Germany}

\author{Martina Hentschel}
\affiliation{Max-Planck-Institut f\"ur Physik komplexer Systeme, N\"othnitzer Str. 38, 01187 Dresden, Germany}

%

%%%%%%%%%%%%%%%%%%%%%%%%%%%%%%%%%%%%%%%%%%%%%%%%%%%%%%%%%%%%%%%%%%%%

\begin{abstract}
We introduce and investigate billiard systems with an adjusted ray
dynamics that accounts for modifications of the conventional reflection of
rays due to universal wave effects. We show that even small modifications
of the specular reflection law have dramatic consequences on the phase
space of classical billiards. These include the creation of regions of
non-Hamiltonian dynamics, the breakdown of symmetries, and changes in the
stability and morphology of periodic orbits. Focusing on optical
microcavities, we show that our adjusted dynamics provides the missing
ray counterpart to previously observed wave phenomena and we describe how
to observe its signatures in experiments. Our findings also apply
to acoustic and ultrasound waves and are important in all situations
where wavelengths are comparable to system sizes, an increasingly likely
situation considering the systematic reduction of the size of electronic
and photonic devices.
\end{abstract}

%%%%%%%%%%%%%%%%%%%%%%%%%%%%%%%%%%%%%%%%%%%%%%%%%%%%%%%%%%%%%%%%%%%%

%


\maketitle

%
%{\bf 

Quantum-classical and ray-wave correspondence are fundamental in 
achieving a joint understanding of the quantum (wave) and classical (ray) 
worlds. In recent years, these ideas have been brought forward 
especially in the field of quantum chaos \cite{stoeckmann} of 
electronic \cite{harayama_book} and optical systems \cite{noeckel}. Besides this 
fundamental, there is a practical aspect: If true, the handy ray picture readily 
provides information about the {\em wave} propagation in optical cavities, 
at seismic interfaces, or in acoustics. Indeed, ray-wave correspondence
was confirmed in numerous situations \cite{noeckel,lee-scar,gmachl-scar,swkim,taka,lebental,dietz}, even away from the formal classical limit of vanishing wavelength~$\lambda$.  In practice, ray-wave correspondence is usually taken for granted. 
However, examples of its violation are accumulating recently, especially in open systems where $\lambda$ is comparable to the 
system size $L$ or radii of curvature $R$ associated with bent interfaces or obstacles.
For instance, in optical microcavities of spiral shape, ``quasiscar''-resonances 
have been computed for which no ray analogue could be identified
\cite{koreaner_spiral}.
Often, Husimi projections \cite{husimi} 
systematically show less symmetry and small mismatches in comparison to the 
classical Poincar\'e surface of section~\cite{lee,PREannbill}.
Note that discrepancies between ray and wave pictures may arise for  {\em all} 
types of waves when $R,L \gtrapprox \lambda$.

\begin{figure}[tb]
\includegraphics[width=\columnwidth]{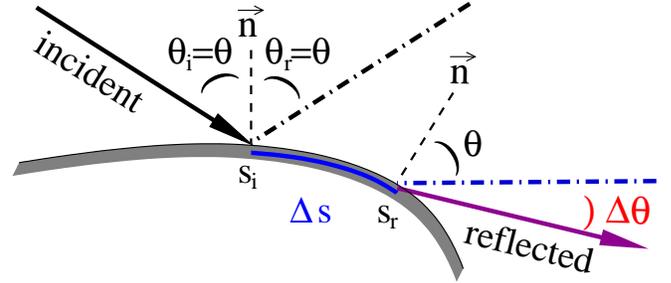}
\caption{(Color online) Generalized reflection law that includes a shift $\Delta s$ along the boundary coordinate $s$ (GHs) and an increase in the angle of the reflected ray from $\theta$ to $\theta + \Delta \theta$ (Ff). The arrows denote the beam center.}
\label{fig1}
\end{figure}

In this Letter, we show that these discrepancies arise when ignoring deviations from the conventional ray picture occurring towards the wave regime. 
We incorporate these deviations in an adjusted reflection law, schematically illustrated in Fig.~\ref{fig1}, and 
we show the dramatic consequences to the long time dynamics of billiards. 
Most remarkably, the ray dynamics typically becomes {\em non-Hamiltonian} and, e.g., attractors and repellors replace the so-called
Kolmogorov-Arnold-Moser (KAM) islands of stability. We show how our adjusted ray dynamics explains the previous observations of quasiscar-resonances and of broken symmetries, and we describe how to test our predictions in the laboratory.

When a light ray hits a glass-air interface, it will split into two rays, one reflected and one refracted, unless the angle of incidence $\theta_i$ (measured to the boundary normal) is larger than a critical angle, $\theta_c = \arcsin 1/n$ (where $n$ is the refractive index of the material, $n=1.5$ for glass) and total internal reflection occurs. Reflection and refraction follow the well-known Snell and Fresnel laws; in particular, the reflection angle $\theta_r$
will equal the incident angle $\theta_i$.
This is the well-established ray picture of geometric optics. In reality, however, the incident light ray will be a {\em beam}, a bundle of rays
with angles of incidence $\theta_b$ centered around a mean angle $\theta_i$. When $\theta_i \approx \theta_c$ 
the larger $\theta_b$ experience total internal reflection, whereas the smaller $\theta_b$ are refracted. Consequently, the mean angle of the reflected beam will be {\em larger} than the original angle of incidence, $\theta_r > \theta_i$: The ray picture is violated! This effect has been termed {\em Fresnel filtering} (Ff) \cite{ff,rex,schomerus} and
was recently measured with microwaves \cite{ff_GH_micro}. Remarkably, a similar effect was observed in chemical reactions in excitable media~\cite{ff_chem}.

%%%%%%%%%%%%%%%%%%%%%%%%%%%%%%%

\begin{figure*}[tb]
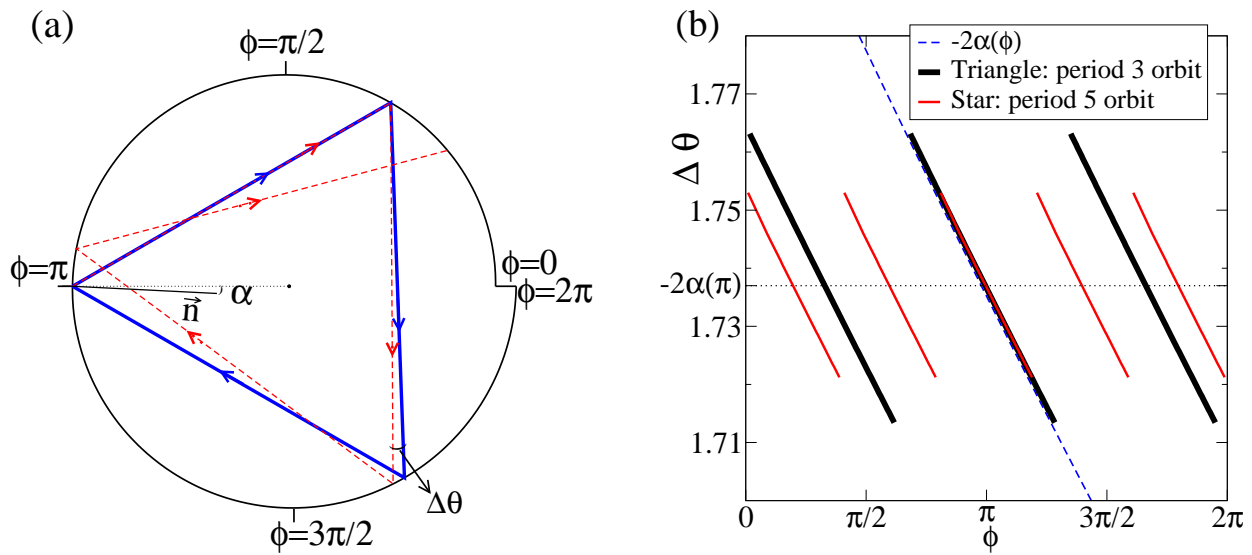

\includegraphics[width=0.9\columnwidth]{fig2a.eps}\hspace{0.1\columnwidth}\includegraphics[width=0.9\columnwidth]{fig2b.eps}
\caption{(Color online) (a) Spiral billiard with $\epsilon=0.1$: conventional (dashed line) and adjusted (solid line) ray dynamics with constant Ff~$\Delta \theta=1.737^\circ$. (b) Values  $(\phi,\Delta \theta)$ for which a period $3$ triangular orbit (black line) and a period $5$ star orbit (red/gray line) exist. The dashed line corresponds to~$2\alpha$, where $\alpha$ shown in (a) is the small angle between the normal and the vector pointing to $R=0$, given by $\sin (\alpha)=\epsilon/\sqrt{\epsilon^2+(1+\epsilon\phi)^2}$.
  }
\label{fig2}
\end{figure*}

%%%%%%%%%%%%%%%%%%%%%%%%%%%%%%%%

Another feature inherent to total internal reflection of beams, also shown in Fig.~\ref{fig1}, is the {\em Goos-H\"anchen shift} (GHs), discovered in 1947 for light but also well-known in acoustics  \cite{goosh,goosh2}.  
Its origin is the interference of rays with slightly different angles of incidence that can be interpreted as a {\em lateral shift} $\Delta s$ of the reflected beam along the material interface, parametrized by~$s$, the arclength of the interface. As both Ff and GHs have the same origin (generic beams consisting of a distribution of $\theta_b$'s) they arise together \cite{schomerus,ff_GH_micro}, as shown in Fig.~\ref{fig1}, but act in complementary phase space directions.
It remains an important open problem to obtain universal analytical expressions of the GHs and Ff at curved interfaces.
For reflections off circular boundaries, the GH and Ff effects were computed in~\cite{schomerus} (see Fig.~\ref{fig3} below), where 
%At this point, a few words on the actual size of Ff and GHs are in order. 
the {\em single} reflection of a Gaussian beam at a circular air hole scatterer in a glass matrix was carefully investigated. 
For a size parameter $n k R=400$ ($k = 2 \pi/\lambda$), corresponding to the semiclassical regime between the pure wave and ray limits, Ff and GHs amount to a few degrees. They are even larger at smaller $nkR$ and appreciable especially near $\theta_c$. 
In this paper, we adopt the terminology used in optics and refer to the corrections leading to non-specular reflection as Ff and GHs, but
emphasize the potential importance of corresponding effects in the reflection of acoustic, seismic \cite{seismic}, or even shock \cite{shock} and chemical~\cite{ff_chem} waves. 

Interestingly, the combined effect of Ff and GHs have never been incorporated 
in the ray dynamics of {\em billiards}, despite their popularity as model systems for
example in the field of quantum chaos.  Here we fill this gap and use the results obtained in

Ref. \cite{schomerus} for a {\it single} reflection to investigate the long-term dynamics in billiards.
 Changes in the phase space are expected~\cite{newref,scars5}, 
not least because of the results obtained for soft-wall billiards \cite{softwall1,softwall2}.
%(similar to GHs). %of the GHs to reflection at soft walls.
We shall see in the following that taking into account Ff and GHs will indeed tremendously
affect  phase space of optical microcavities that are described with the adjusted ray
optics. In particular, we will find that the emerging new features are in much better
agreement with previously obtained wave dynamical results -- using the adjusted ray optics profoundly
deepens our understanding of the wave dynamics in open (optical) quantum-chaotic systems.

%%%%%%%%%%%%%%%%%%%%%%%%%%%%%%%%%%%%%%%%%%%%%%%%%%%%%%%%%%%%%%%%%%%% SPIRAL %%%%%%%%%%%%%%%%%%%%%%%%%%%%%%%%%%%%%%%%%%%%%%%%%%%%%%%%%%%%%%

\begin{figure*}[tb]
\includegraphics[width=1.8\columnwidth]{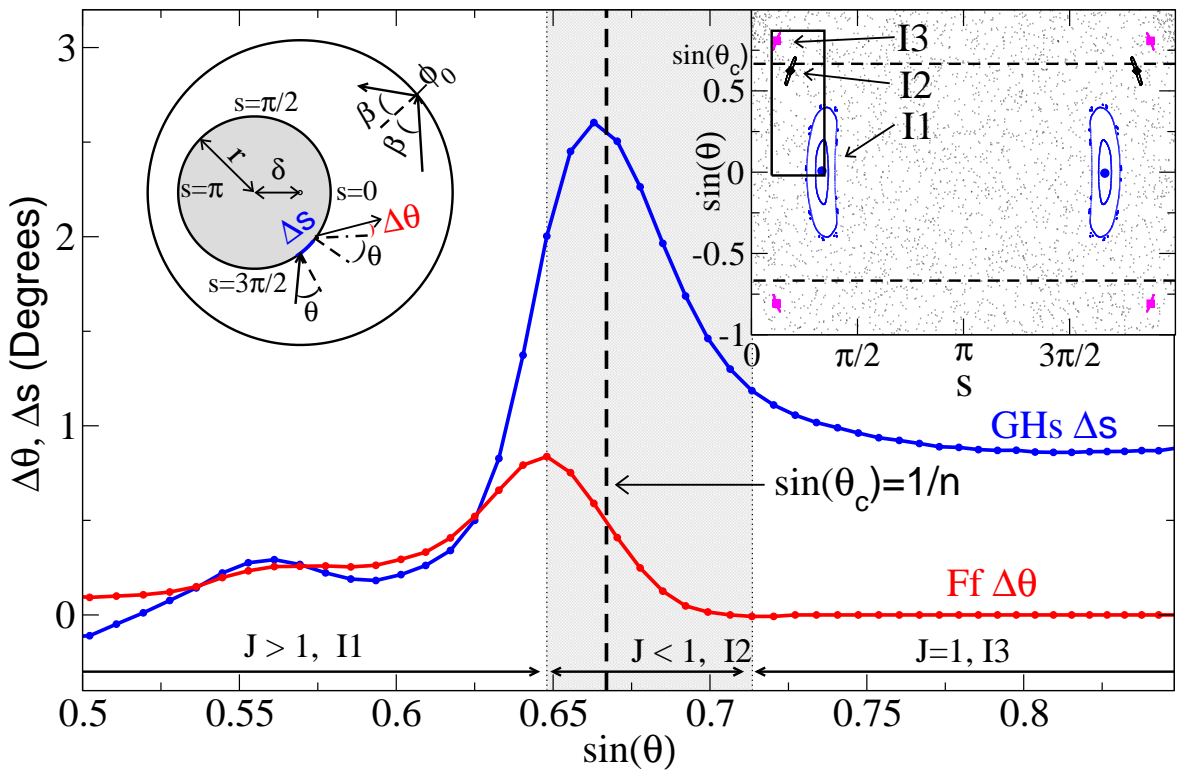}
\caption{(Color online) Terms of the adjusted reflection law: GHs~$\Delta s$ (upper curve)
  and Ff~$\Delta \theta$ (lower curve) for a circular obstacle with $n=1.5$ and $kR=400$
  (from~\cite{schomerus}). The correction is largest near the critical angle~$\theta_c$;
  we assume a monotonic increase of Ff towards $\theta_c$. Insets: Annular billiard
  geometry (left) and its conventional phase space (right), obtained through a Poincar\'e
  section ($\sin \theta$ vs.~spatial position $s$) at the {\em inner} boundary. 
  }\label{fig3}
\end{figure*}
%%%%%%%%%%%%%%%%%%%%%%%%%%%%%%%%

We first consider the effect of the adjusted ray dynamics in spiral billiards defined by $R(\phi)=1+\epsilon \phi/2\pi$, in polar coordinates, as shown in Fig.~\ref{fig2}(a).
We focus on rays rotating in the clockwise direction, as the one denoted by the dashed line, because the other rays quickly hit the ``notch'' at $\phi=0$ and leave the cavity. 
For these rays, the angular momentum~$M$ (being the shortest distance from the ray to the center of the spiral) 
decreases with every collision due to the small angle $\alpha$, described in Fig.~\ref{fig2}.
This implies that no periodic orbit avoiding the notch exists. Ray-wave correspondence
suggests that the same should be true for (wave) resonances. Therefore, the finding of
triangular ($n=2$) and star-shaped ($n=3$) resonances in Ref. \cite{koreaner_spiral} which
were named ``quasiscars'' came as a true surprise.

Let us now correct the conventional ray picture by adding (constant) Ff and GHs.
The Ff increases the reflection angle by $\Delta \theta$,
which counteracts the above-mentioned decrease of $M$. The GHs~$\Delta \phi$ increases $M$, however, this effect is roughly $\epsilon/2\pi$ times smaller than Ff (for $\Delta \theta \approx \Delta \phi$), and will be neglected~\cite{note_ghs}.
For a full interval of values $\Delta \theta$ (around $2\alpha$) the conservation of $M$ can be fully restored after several collisions, allowing periodic orbits to exist. 
The solid curve in Fig.~\ref{fig2}(a) shows one of such %resulting
(unstable) {\em periodic} orbits, obtained for $\Delta \theta = 1.737^\circ (=2\alpha$ for $\epsilon=0.1$ and $\phi=\pi$).
%\cite{footnote_ff} 
As discussed above and shown in Fig.~\ref{fig3}, this value is consistent with the value of Ff expected for the parameters used in Ref.~\cite{koreaner_spiral} ($\theta \approx \theta_c$ and $nkR \approx 100$).
Figure \ref{fig2}(b) shows that changing $\Delta \theta$ yields triangular orbits slightly rotated in space (an effect also seen in Ref.~\cite{koreaner_spiral}); 
in fact, for almost all reflection points  $\phi$ there is a $\Delta \theta$ for which a periodic orbit exists. We found equivalent results for star shaped orbits and other periodic orbits with small period. %\cite{we_in_prep}. 
The authors of Ref.~\cite{koreaner_spiral} correctly argued the closeness of $\theta_i$ of the regular resonances to $\theta_c$ to be crucial. We have now clarified 
that Ff provides the explanation -- note that GHs alone does not allow the angular momentum $M$ to retain its initial value along the orbit -- and that the regular trajectories seen are thus {\em real} scars \cite{heller} of a spiral billiard with adjusted ray dynamics. The above results lead to the following predictions for the experiments that are being performed: $\Delta \theta(k)$ decreases as $k r$ increases, and so quasiscars are expected to appear at different locations~$\phi$ until Ff becomes too small and no quasiscars are predicted to appear, in agreement with the semiclassical limit.  We predict also that depending on~$n$ other orbits will be stabilized, e.g., the square orbit for $n \approx \sqrt{2}$.

%%%%%%%%%%%%%%%%%%%%%%%%%%%%%%%

\begin{figure*}[tb]
\includegraphics[width=1.8\columnwidth]{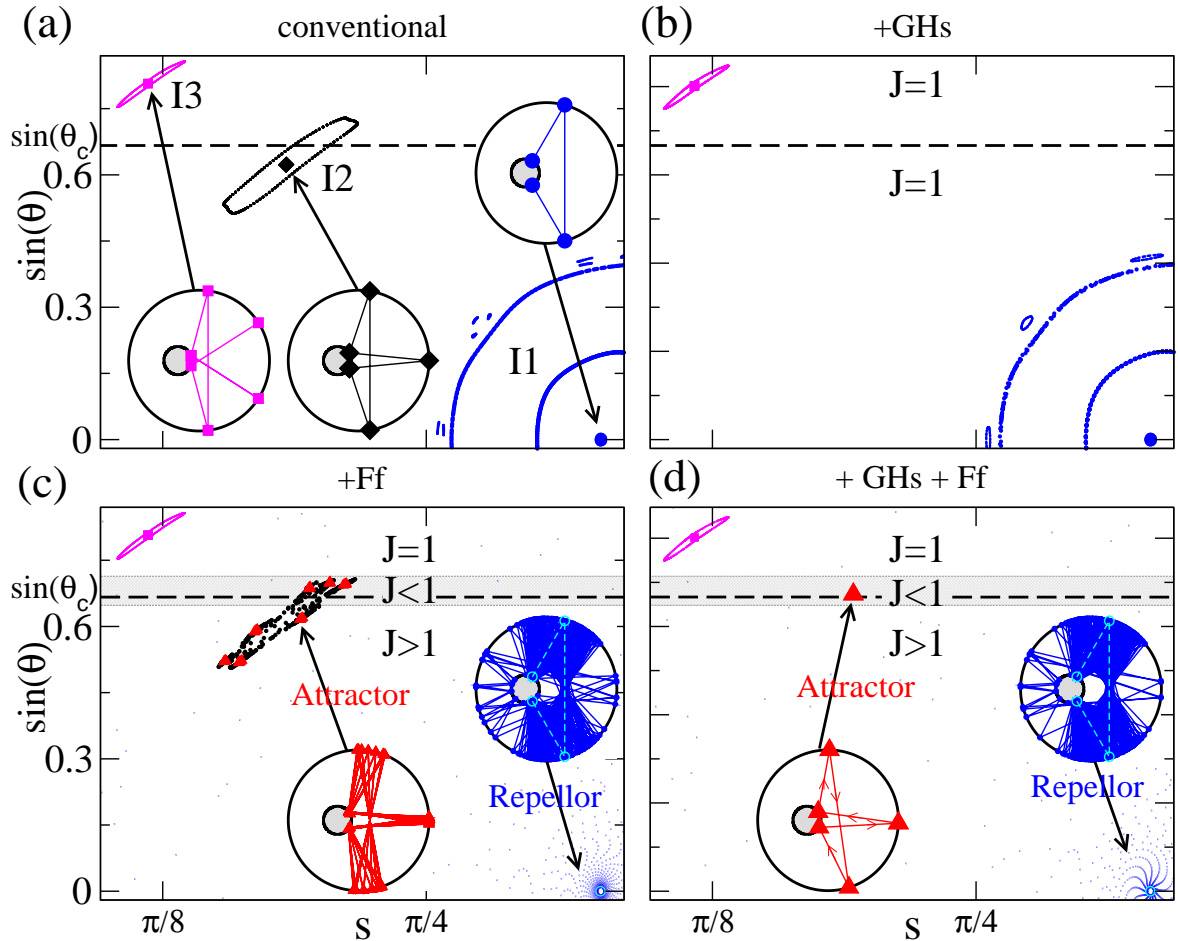}
\caption{(Color online)
Phase space of the annular billiard~\cite{annular} with (a) conventional reflection law (magnification of right inset of Fig.~\ref{fig3}), and including (b) GHs, (c) Ff, and (d) GHs and Ff, as described in Fig.~\ref{fig3}. Note that GHs modifies KAM islands in (b), but preserves the Hamiltonian character  that is lost when
Ff is taken into account in (c) and (d), where attractors ($\blacktriangle$)       
and repellors ($\circ$) are formed. Insets show the corresponding orbits in the real space. %See text for details.
  }\label{fig4}
\end{figure*}

%%%%%%%%%%%%%%%%%%%%%%%%%%%%%%%%

Let us now investigate in more detail the joint effect of Ff and GHs on the phase-space dynamics ($s, \sin \theta $), that is most often used to establish the ray-wave correspondence. 
We adjust the conventional reflection law by replacing

\begin{equation}	\label{eq.trans}
	\left(
	\begin{array}{cc} s_r \\ \theta_r \end{array}
	\right)
	=
	\left(
	\begin{array}{cc} s_i \\ \theta_i \end{array}
  \right)
%  \longrightarrow
%\text{ by }
\rightsquigarrow
  \left(
  \begin{array}{cc} s_r \\ \theta_r \end{array}
	\right)
	=
	\left(
	\begin{array}{cc} s_i + \Delta s(s_i,\theta_i) \\ \theta_i + \Delta \theta(s_i,\theta_i) \end{array}
	\right) \:.
\end{equation}

The dynamics in a billiard is described by the billiard transformation~$T$ that maps a collision 
$(s_i, \sin \theta_i)$ to the next one. Considering~(\ref{eq.trans}), the absolute value of the Jacobian of $T$ is given by 
\begin{eqnarray}
 %\label{eqJ}
J & = & |\frac{\cos(\theta+\Delta \theta)}{\cos(\theta)} \times \label{eqJ} \\
& & 
\left(1+\frac{\partial \Delta\theta}{\partial \theta}+ \right.
\left. \frac{\partial \Delta s}{\partial s}+ \right.
\left. \frac{\partial \Delta  \theta}{\partial \theta}\frac{\partial \Delta s}{\partial s} \right.
\left. -\frac{\partial  \Delta \theta}{\partial s}\frac{\partial \Delta s}{\partial\theta}\right)|\;. \nonumber
%\end{array}
\end{eqnarray}
%
%%%%%%%%%%%%%%%%%%%%%%%%%%%%%%
%
To  understand the new billiard dynamics is crucial to know whether $J$ deviates from~$1$.
The phase-space area $\cos \theta d s d \theta$ 
is preserved if and only if $J=1$, which, in two dimensions, implies Hamiltonian dynamics (evidently, $J=1$ for 
$\Delta s = \Delta \theta =0$). For $J \neq 1$, the map does not preserve the area $\cos \theta d s d \theta$ and typically
non-Hamiltonian dynamics arises.
Equation (\ref{eqJ}) indicates that $J$ sensitively depends on the system-specific 
dependence of $\Delta \theta$ and $\Delta s$ on $\theta$ and $s$,
allowing, in principle, $J$ to take arbitrary values.

We are interested on how the modification~(\ref{eq.trans}) affects the dynamics of a billiard for which the conventional reflection law produces a generic phase space, i.e., coexistence of chaotic and regular dynamics. A long studied~\cite{annular} and physically realistic example is the annular billiard shown in the left inset of Fig.~\ref{fig3}: A glass disk with conventional reflection law at the outer boundary (realizable by a mirror coating) and an eccentrically placed inner disk scatterer (of radius $r$ and distance $\delta$ between the disk centers). Ff and GHs act thus only at the inner boundary, where the Poincar\'e section is defined. For GHs and Ff we use the values reported in Ref.~\cite{schomerus}, cf. Fig.~\ref{fig3}. We are not interested in the refractive intensity loss at the inner boundary, that could be included via Fresnel's law \cite{PRERapid}. Rather, our focus is on the dynamics of rays that remain in the annulus.

In this system, the curvature is constant, $R=r$, and Eq.~(\ref{eqJ}) simplifies to
\begin{equation}\label{eqJannular}
J = \frac{\cos(\theta+\Delta \theta)}{\cos(\theta)}  
\left(1+\frac{\partial \Delta\theta}{\partial \theta} \right)\;.
\end{equation}
Because $\Delta \theta$ is a small correction, mainly the last term
$\partial \Delta \theta/ \partial \theta$ determines $J$. Three regions, with $J>1, J<1$, and $J=1$, are marked in Fig.~\ref{fig3}.
Next we compare 
%consider differences in 
the phase space of billiards with conventional dynamics and those with GHs and Ff corrections. %, and their joint effect. % (GHs and Ff are taken into account for reflections at the inner disk). 
We have chosen parameters $r=0.2, \delta=0.3$ in the annular billiard such that at least one KAM island is seen % CHECK: ZITAT 
in each of the three regions, named I1, I2, and I3 and indicated in the right inset of Fig.~\ref{fig3}. The boxed region is enlarged in Fig. \ref{fig4}a, where the stable periodic orbits in the center of the islands are shown as insets. The following panels (b)-(d) show that the effect of GHs and Ff can destroy, create, and modify these KAM islands. Now, we explain these observations.

The change in the structure of the phase space is related to the value of $J$, cf. Eq.~(\ref{eqJannular}) and ~Fig.~\ref{fig4}.
While elliptic islands contained in the area-preserving region $J=1$ ($\theta \gg \theta_c$, I1) survive when GHs and Ff are switched on,
those contained in the region $J<1$ ($J>1$), where the dynamics is area contracting (expanding), become attractors (repellors) once the Ff correction is taken into account.
The area contracting ($J<1$) region is 
placed around the critical angle $\theta \approx \theta_c$ and affects I2 that gives rise to a periodic attractor in Fig.~\ref{fig4}(c)-(d).

Note that although this (effective) dissipation is related to the openness of the system, 
via the partial refractive escape of light that gives rise to the Ff effect,
% but that we are not interested in Fresnel intensities along the orbits but rather in the % % dynamical structure of the phase-space.
%but that the possibility of
the refractive escape itself is {\em not} its origin. % of dissipation. 
This becomes evident as also $J>1$ is realized for subcritical incidence $\theta <
\theta_c$, and I3 originates a repellors in Fig.~\ref{fig4}(c)-(d).
This dissipative behavior is related to the openness of the system via the partial refractive escape of light that gives rise to the Ff effect. 
Note, however, that the possibility of refractive escape and the Ff effect may also generate {\it expanding} dynamics ($J>1$), i.e., the first-sight impression that refractive escape can only lead to contracting dynamics because of the loss of energy through transmitted rays
does not apply. (Note also that we are not interested in the refractive intensity losses in the present study.)  Whether the Jacobian $J$ takes a value larger or smaller than 1 depends, in the case of the annular billiard, cf.~Eq.~(\ref{eqJannular}), on $\partial \Delta \theta / \partial \theta$, i.e., on the dependence of the Ff correction $\Delta \theta (\theta)$ on the angle of incidence $\theta$ [see Eq.~(\ref{eqJ}) for the general case].
Indeed, for subcritical incidence $\theta< \theta_c$ we find $ \partial \Delta \theta
/ \partial \theta > 0$ and Eq.~(3) leads to  $J>1$. Accordingly, the KAM island I3,
present in the region of  subcritical incidence in Fig. 4(a), transforms into a repellor
when Ff is considered in Fig. 4(c)-(d).
The action of GHs alone, cf. Eq.~(\ref{eqJannular}) and~Fig.~\ref{fig4}b, preserves the
Hamiltonian dynamics ($J=1$). In this case, modifications of the phase space similar to
those obtained by changing a control parameter occur, e.g., in Fig.~\ref{fig4}, region I2 is absent and small islands are created in the secondary chain of I1 and I3 (hardly visible). Most exciting and dramatic in its effect is the Ff contribution that renders islands into attractors (close to~$\theta_c$) and repellors, i.e., signatures of a {\em non-Hamiltonian} dynamics ($J\neq1$) are seen in a model system often used in quantum and wave chaos.

Another important effect of GHs and Ff is the breaking of symmetries.
The conventional phase space of billiards with one symmetry axis, like the annular billiard, possesses
the time reversal symmetry $I_t : (s,\sin\theta) \rightarrow (s, -\sin \theta)$
as well as a spatial symmetry $I_s : (s,\sin\theta) \rightarrow (2 \pi -s, -\sin \theta)$.
In the example above, Ff violates the $I_t$-symmetry.
The simplest way to see this is to consider a beam with (average) critical incidence. The
subcritical part of the beam will be refracted, and the remainder will be reflected to
supercritical angles, violating the rules of specular reflection. If this reflected part
of the beam is now considered as the (time-reversal partner of the original) incident
beam, it becomes immediately evident that time-reversal symmetry and the principle of
ray-path reversibility are lost as all rays of this beam will be totally reflected and
specular reflection is obeyed.  
As a consequence of the violation of time-reversal symmetry due to Fresnel filtering, the $I_t$ symmetric orbit in the center of I2 shown in Fig.~\ref{fig4}(a) becomes Ff- and GHs-distorted, see the left inset in Fig.~\ref{fig4}(d). 
Moreover, for the opposite sense of traveling ($\sin \theta < 0$), the distortion will be in the opposite direction. This shows that the position of the reflection point at the outer boundary depends on the sign of $\theta$, and that 
$I_t$-symmetry is broken. In wave calculations for optical microcavities (where Ff and GHs are naturally included), indications of $I_t$-breaking 
in the Husimi projection \cite{husimi} were noticed~\cite{PREannbill,PRAunidirWG},  but their origin remained unclear.
The adjusted reflection law that we propose here obviously accounts for this symmetry
loss.

The consequences of an adjusted, non-specular reflection law are                                                                                   
accessible to experiments using, e.g., state of the art microwave                                                                                
resonators with the geometry adjusted such that periodic orbits (KAM                                                                               
islands) exist close to the critical line. For example, the annular geometry                                                                               
suggested here can be realized as a teflon disk with an air hole and an                                                                            
outer metallic coating. One approach is the analysis of the length                                                                                
spectra, taken as Fourier transform of the resonances in a finite                                                                                  
interval of wavenumbers $k$. 
For different $k$ intervals, the resonant wave patterns adjust to the new $k R$-dependent Ff and GHs. Therefore we expect changes in the peak position (length of the adjusted orbit) and peak height (stability of the adjusted orbit or size of KAM island).
Another possibility is to directly observe the                                                                                       
$k$-dependent morphology of resonances by directly measuring their wave                                                                            
patterns. % as is indeed possible with nowadays techniques}~\cite{microwaves}.                                                                                        

In principle, our results also apply to ballistic electronic mesoscopic systems such as
quantum dot billiards \cite{qudotbill} since the Fermi wavelength of the electrons is
lower, but comparable to the system size. In addition, the confinement of the electrons in
the two-dimensional electron gas layer is by gate electrodes and will never be perfect. In
this sense the confining potential has soft walls, and corrections to the hard-wall ray
dynamics are expected. We point out that most works in electronic mesoscopic systems focus
on transport properties. Such an openness of the system (by attaching leads and applying
bias voltages) is, however, very different from the openness that we consider here.

In conclusion, we have investigated billiards with an adjusted ray dynamics that incorporates wave corrections to the law of specular reflection. We have shown that the conventional phase space experiences dramatic changes, including the creation of periodic orbits, the breakdown of symmetries, and the creation of regions of non-Hamiltonian dynamics where attractors and repellors replace KAM islands. We have presented strong evidences that these results explain previous observations of 'quasi-scared' modes in spiral microcavities and of broken time reversible symmetry. We expect that other violations of the ray-wave correspondence can be explained similarly, e.g., the observation of disproportionally many regular, scarred, and scar-like resonances~\cite{lee-scar,gmachl-scar,koreaner_spiral,lee,scars,scars2,scars3,scars4,scars5}  might be related to the creation of periodic attractors, and discrepancies in the far-field pattern~\cite{rex,lee} might be explained by the effect of Ff and GHs on invariant sets and their unstable manifolds. 
Although we focused on the case of optical microcavities, where Ff and GHs are well established effects, our main results apply likewise to acoustic, seismic, and ultrasound waves whenever deviations from the specular
reflection become relevant, e.g., when the wave length becomes comparable to the system size.  We want to emphasize the importance of previous unexplained disagreements between ray and wave results, including those potentially regarded as numerical or experimental errors, 
because they might be explained by the adapted ray dynamics introduced here.

\acknowledgments
We thank S. Bittner, B. Dietz, H. Kantz, U. Kuhl, A. Richter, and H.-J.
St\"ockmann for useful discussions. M.H. thanks the DFG for support in
the Emmy-Noether Programme and in the research group FG 760.


\begin{thebibliography}{99}
\bibitem{note_ghs} If GHs is considered, a small increase of the Ff value is enough to reestablish the conservation of $M$.
\end{thebibliography}
\end{document}